\documentstyle[eqsecnum,aps,twocolumn]{revtex}

\newtheorem{lemma}{Lemma}
\newtheorem{definition}[lemma]{Definition}
\newtheorem{theorem}[lemma]{Theorem}

\newtheorem{remark}[lemma]{Remark}
\newtheorem{corollary}[lemma]{Corollary}

\newenvironment{proof}{{\bf Proof:}}{\hfill$\Box$}
\begin{document}
\draft
\preprint{HEP/123-qed}
\title{More Robust Multiparty Protocols with Oblivious Transfer}
\author{J. M{\"u}ller-Quade and H. Imai}
\address{Imai Laboratory, Institute of Industrial Science, The University of 
Tokyo}
\date{May $2^{nd}$, 2001}
\maketitle

%
%

\begin{abstract}
With oblivious transfer multiparty protocols become possible even in
the presence of a faulty majority. But all known protocols can be
aborted by just one disruptor.

This paper presents more robust solutions for multiparty protocols
with oblivious transfer. This additional robustness against disruptors
weakens the security of the protocol and the guarantee that the result
is correct. We can observe a trade
off between robustness against disruption and security and
correctness. 

We give an application to quantum multiparty protocols. These allow
the implementation of oblivious transfer and the protocols of this
paper relative to temporary assumptions, i.e., the security increases
after the termination of the protocol.
\end{abstract}

%
%

\section{Introduction}


In a multiparty protocol a set $P$ of players wants to correctly
compute a function $f(a_1,\dots,a_n)$ which depends on secret inputs
of $n$ players. Some players might collude to cheat in the protocol as
to obtain information about secret inputs of the other players or
to modify the result of the computation.
  
Using oblivious transfer multiparty protocols for arbitrary functions
can be carried out with unconditional security if all players are willing
to cooperate~\cite{BeaGol89,GolLev90,CreGraTap95}. But already one
disruptor can abort the protocol without being identified.


This contribution generalizes the unconditionally secure multiparty
protocols with oblivious transfer to the case where we don't only care
about security and correctness but also about robustness against disruptors.

The basic idea is to define protocols where either no possible
collusion of disruptors can abort the protocol or a cheater is
identified unambiguously. Then the protocol can be restarted without
the cheater.

To be able to enhance the robustness of protocols we have to make some
assumptions about possible collusions. We model possible
collusions by defining a set of collusions. Only one of these possible
collusions is actually cheating. Within this set of colluding
players the players share their input and take actions based on their
common knowledge.

\begin{definition}
An adversary structure is
a monotone set ${\cal A}\subseteq 2^P$, i.\,e., for subsets
$S'\subseteq S$ of $P$ the property $S\in {\cal A}$ implies $S' \in
{\cal
  A}$. By assumption one set of $\cal A$ contains all cheaters.
\end{definition}

The main properties of a multiparty protocol are:
\begin{enumerate}
\item A multiparty protocol is said to be ${\cal A}$-{\em secure} if
no single collusion from $\cal A$ is able to obtain information about
the secret inputs of other participants which cannot be derived from
the result and the inputs of the colluding players.
\item A multiparty protocol is ${\cal A}$-{\em partially correct} if no
possible collusion
  can let the protocol terminate with a wrong result.
\item A multiparty protocol is ${\cal A}$-{\em correct} whenever no
  single collusion from $\cal A$ can abort the protocol, modify its
  result, or deviate from the protocol in a way that an honest player
  obtains information about the secret inputs of another player which
  cannot be derived from the result and the input of this honest player.
\item A multiparty protocol is called $\cal A$-{\em fair} if no
  collusion from $\cal A$ can reconstruct the result of the multi
  party computation earlier then all honest participants together. No
  collusion should be able to run off with the result.
\end{enumerate}

A multiparty protocol having the properties 1., 2. and 4. is called
${\cal A}$-{\em partially robust} and a protocol having
all four above properties is called
${\cal A}$-{\em robust}.

By assumption one set of $\cal A$ contains all players who deviate
from the protocol.
Hence we demand security against only one possible collusion which
has to contain all cheaters as well as all disruptors.

Sometimes a set $M$ is able to reconstruct a secret due to the
cheating of players who are not contained in $M$. In this situation we
will always look at the complete set of cheaters necessary to obtain
this situation. So even in an
$\cal A$-secure protocol with $M\in{\cal A}$ it can happen that the
players of $M$ can reconstruct a secret if some players outside of $M$
are cheating, too. 

All players are considered to be curious, i.\,e., even the honest
players will try to learn as much as possible from the information
accessible while following the protocol.

%
%

\section{Previous Work}\label{secretSharing}

\subsection{Multiparty Computations with Private Channels}

We will summarize next what can be achieved by classical multiparty
computations when private channels are available between any two
players as well as a broadcast channel. The next result is
taken from~\cite{HirMau97} for a history see references therein.

\begin{theorem}
  Given a set $P$ of players with an authenticated secure channel
  between each pair of players together with a
  broadcast cannel, then every function can be computed by an $\cal
  A$-partially robust multiparty protocol if no two sets from $\cal A$
 cover
  the complete set $P$ of players.
\end{theorem}

But if we additionally conslider disruption the result does not hold
any more:

\begin{remark}
  There exist functions for which a multiparty protocol among players
who have access to a broadcast channel and have authenticated secure
channels connecting every pair of players cannot be $\cal A$-robust if
two collusions cover $P\setminus \{ P_i\}$ for some player $P_i$.
\end{remark}

\begin{proof}
Whenever two players mutually accuse each other to not use the secure
channels approriately it is not possible for the remaining honest
players to decide which of the two players is cheating and the secure
channel between the two players cannot be used.

If the players of two possible collusions $A_1, A_2\in {\cal A}$
covering $P\setminus \{P_i\}$ cannot use the secure channels between them
for the above reason, then again it is not clear for $P_i$ which of
the two possible collusion is cheating. To continue with the protocol
all messages between players who are complaining about each other have
to be exchanged over the broadcast channel or over secure channels via
$P_i$. Obviously $P_i$ learns all secrets or the protocol must be
aborted. In both cases the protocol is not $\cal A$-robust.
\end{proof}


\subsection{Multiparty Computations with Oblivious Transfer}

This subsection summarizes the previous work on multiparty protocols
with oblivious transfer.

In the following we will always think of the oblivious transfer
channel as being a stronger primitive than authenticated private
channels. All oblivious tranfer channels in the remainder of this
paper are authenticated and secure and we will not state
these properties any more.


Given an oblivious transfer channel all secure two party computations
become possible with perfect security~\cite{Kil88}. This result was
generalized to allow multiparty computations with a dishonest
majority~\cite{BeaGol89,GolLev90,CreGraTap95}. One obvious problem with
such
protocols is that if a majority of players cannot run off with the
secret, i.\,e., they cannot reconstruct the secret on their own, then
a minority of players can abort the protocol.
This is problematic if it is not clear who is cheating otherwise the
protocol could be restarted without the cheaters.
To capture what can be achieved in this case we
made the distinction between partial correctness and correctness.

The result of~\cite{BeaGol89,GolLev90,CreGraTap95} can then be stated as

\begin{theorem}
  Given an oblivious transfer channel between any two
  players as well as a broadcast channel then every
  function can be realized by a $\emptyset$-robust, $2^P$-secure,
  $2^P$-fair, and
  $2^P$-partially correct multiparty protocol.
\end{theorem}

\section{More Robust Multiparty Protocols: An Overview}

In this section we shortly overview the protocol presented in this
paper to simplify reading. We shortly explain the basic primitives
of~\cite{CreGraTap95} and sketch our changes to the protocol to obtain
more robustness. All these changes will be carried out in detail in
the following sections.

As long as no conflict occurs we will follow the protocol
of~\cite{CreGraTap95}, but whenever disruption takes place we
will deviate from the protocol in a way that either resolves the
problem or a cheater is identified. Then the protocol can be restarted
without the cheater. The exclusion of a player has an impact on the
value of the function to be computed. The best way to deal with this
would be to have a default input like ``unvalid'' in a voting
scheme. But the effect of the exclusion of such a player is not severe as
the player was a cheater anyway and could as well have chosen a random
input.  We will not discuss this further as the discussion depends
strongly on the function to be implemented.  There is one more
problem with restarting a protocol. If the inputs were time critical
someone might force a restart just to be able to change his inputs,
this will be avoided in Remark~\ref{Restart}.

To be able to restart the protocol without players who did try do
disrupt the protocol we have to be able to identify
cheaters. To do this we will replace the subprotocols used
in~\cite{CreGraTap95} by primitives which either terminate
successfully or a cheater is identified.

Following~\cite{CreGraTap95} we will first give a bit commitment
protocol which binds a player to all other players and allows for zero
knowledge protocols of linear relations on committed bits. In
Section~\ref{SectionGBCX} we will introduce a protocol which either
successfully creates such a bit commitment or a cheater is identified.

Based on this bit commitment one can generate distributed bit
commitments where a bit is shared among all players and each player is
committed to his ``share''. Section~\ref{SectionDBC} gives a protocol which
either successfully creates such a distributed bit commitment or a
cheater is identified.

In~\cite{CreGraTap95} a {\em committed oblivious transfer} protocol
was introduced which allows to implement the boolean function AND on
distributed bit commitments. Section~\ref{SectionGCOT} gives a variant of
this protocol which allows to identify a cheater in the case that the
protocol fails. The same techniques allow to implement a NOT function
on a distributed bit commitment.

with these boolean function we can realize every boolean function on
distributed bit commitments by circuit evaluation.

The outline of the complete protocol, given in
Section~\ref{SectionProtocol}, then is:

{\bf Initialization Phase}: All players have to agree on the function
to be computed as well as on the circuit $F$ to be used, they have to
agree on an adversary structure $\cal A$ such that the protocol will
be $\cal A$ robust and all players have to agree on the security
parameters.
Furthermore the players agree on how to, in case of a restart of the
protocol, choose the input of a cheater which has been excluded from
the protocol.

Then all players create distributed bit commitments to commit to their
inputs. All players will be able to generate distributed bit
commitments or a cheater is identified and the protocol can be
restarted without the cheater.

{\bf Computing Phase}: The circuit $F$ is evaluated using AND and NOT
gates on the distributed bit commitments. Each of these boolean
gates is either applied successfully or a cheater is identified and
the protocol can be restarted without him.

{\bf Revelation Phase}: The result of a computation is hidden in the
``shares'' of distributed bit commitments. These have to be unveiled
in a way to ensure the fairness of the
protocol. Following~\cite{CreGraTap95} this can be done by techniques
known in the literature~\cite{Cle89,GolLev90}.

In Section~\ref{SectionAfterTermination} we analyze the situation
after the protocol has terminated. After the termination we can make
more precise statements about the security.

\section{More Robust Multiparty Protocols: What Is  Impossible}
The aim of this contribution is to enhance the robustness of multiparty
protocols with oblivious transfer. No possible collusion should be able
to abort the protocol. This results in a tradeoff between security,
parial correctness, and
robustness which will be analyzed in the following.

We first give a bound on the robustness which can be achieved. The
bound is tight as our protocols reach this bound.

\begin{lemma}\label{MPNoGo}
  Let $P$ be a set of players for which each pair of players is
  connected by an oblivious transfer channel and each player
  has access to a broadcast channel.  Then $\cal
  A$-robust multiparty computations are impossible for all functions
  if two sets of $\cal A$ cover $P\setminus \{ P_i\}$ for
  a player $P_i\in P$ or $|P|=2$.
\end{lemma}

\begin{proof}
Whenever two players mutually accuse each other to not properly use
the oblivious transfer channel it is impossible for the remaining
honest players to decide which of the two players is actually refusing
to cooperate. 
Let $A$ and $B$ be two possible collusions covering $P\setminus \{ P_i
\}$, such that all the players from $A$ are in conflict with all the players
from $B$ about refusing to use the oblivious transfer channel. Then
the oblivious transfer channels between the players of $A$ and 
the players of $B$ cannot be used and it is impossible for $P_i$ to
decide who is cheating. 
 
The player $P_i$ must assist the players from $A$ and $B$. As no other
player can assist we are in the three party situation with an
oblivious transfer channel only between a player Alice and $P_i$ and a
player Bob and
$P_i$. For each bit being transferred from Alice to Bob the player
$P_i$ knows either as much as Alice about this bit or he knows as much
as Bob. The players Alice and Bob cannot agree on a bit known to both
without $P_i$ knowing it, too. Many functions can hence not be
computed by multiparty protocols in this situation.
\end{proof}


\section{The Structure of Conflicts}

In this paper we will often take actions depending on an analysis of
the complaints some players have broadcasted about other players. For
this we introduce the notion of a {\em conflict} and look at the
computational complexity of such an analysis.

\begin{definition}
  We say that two players $P_i,P_j\in P$ are {\em in conflict} with
  each other if all honest players can derive that one of the two
  players is cheating.

  Whenever all honest players can conclude that either all players
  from a set $A\subseteq P$ or all players from a set $B\subseteq P$
  are cheating we say that these two sets are {\em in conflict} with
  each other.
\end{definition}

E.\,g. if a player accuses some other player of cheating these two players
are in conflict as either the first player is lying or the second is
cheating. Every honest player must complain about every player he
knows is cheating. A player who does not report every cheating he
detects is thought to be colluding with the dishonest players.

It is clear from the definition that two sets $A,B$ are in
conflict if and only if every player from $A$ is in conflict with
every player from $B$.

With the set $P$ of players and the conflicts which occured we can define
a {\em graph of conflicts}. Together with the adversary structure
$\cal A$ we will call it the {\em conflict structure}.

\begin{definition}
A graph $\Gamma$ with the set $P$ being the vertices and two vertices
being connected by an edge iff the two players are in conflict is called the 
{\em graph of conflicts}. 

A pair $(\Gamma,{\cal A})$ with $\Gamma = (P,E)$ being a graph of
conflicts and ${\cal A}\subseteq 2^P$ being an adversary structure is
called a {\em conflict structure}.
\end{definition}

To be able to identify possible collusions which could or cannot be
responsible for a given graph of conflicts we define the {\em vertex
cover}.

\begin{definition}
For a graph $\Gamma = (P,E)$ a vertex cover is defined as a subset of
the vertices which contains for every edge at least one vertex
incident with this edge.
\end{definition}

To get results about the complexity of some problems concerning
conflict structures we recall the $t$-{\em vertex cover problem}. The
$t$-{\em vertex cover problem} is the problem to decide for a given
graph if it contains a vertex cover of size $t$ or less. In the
following it is of interest that this decision problem is known
to be ${\cal NP}$-complete~\cite{GarJoh79}.

\begin{remark}\label{Rem}
Let $(\Gamma, {\cal A})$ be a conflict structure and let $\cal C$
denote the set of vertex covers of $\Gamma = (P,E)$. Then the set of
all cheaters is contained in ${\cal C}\cap {\cal A}$. If no vertex cover
of $\Gamma$ is contained in $\cal A$ then the assumption that the set
of all cheaters is contained in $\cal A$ is violated.
\end{remark}

\begin{proof}
By assumption one set of $\cal A$ contains all cheaters, so we have to
show that a set from $\cal A$ which is not a vertex cover cannot
contain all cheaters. This is trivial as for every set $M$ from $\cal
A$ which is not a vertex cover of $\Gamma$ there exists a pair of
players who are in conflict but neither of them is contained in
$M$. By the definition of conflicts one of the two is cheating, but
not contained in $M$ and thus $M$ cannot contain all cheaters.

If no vertex cover of $\Gamma$ is contained in $\cal A$, then for
every set $M$ of $\cal A$ there exists a pair of players who are in
conflict but neither of them is contained in $M$. Hence no set of
$\cal A$ contains all cheaters and the assumption is violated.
\end{proof}

One can view ${\cal C}\cap {\cal A}$ as the updated adversary
structure after taking into account the conflicts present.

The above remark yields a simple, but not efficient, algorithm, to
identify a cheater whenever a cheater can be identified based on the
conflict structure present.

\begin{lemma}
Let $(\Gamma, {\cal A})$ be a conflict structure, let $\cal C$ denote
the set of vertex covers of $\Gamma = (P,E)$, and let $M$ be the set
of all cheaters which can be identified by deduction from the conflict
structure.

Then $$M = \bigcap_{S\in{\cal C}\cap {\cal A}}S.$$
\end{lemma}

\begin{proof}
For every vertex cover $C\in {\cal C}\cap{\cal A}$ it is consistent
with the conflict structure to assume that only the players in $C$ are
cheating.  Hence if there exists a vertex cover $C\in {\cal
C}\cap{\cal A}$ which does not contain a specific player, then this player
need not be a cheater. So whenever a cheater can be identified by
deduction from the conflict structure he must be contained in
$\bigcap_{S\in{\cal C}\cap {\cal A}}S$. On the other hand if this
intersection is not empty then every player in this intersection must
be cheating as one set of ${\cal C}\cap{\cal A}$ contains only
cheaters, which follows from Remark~\ref{Rem}.
\end{proof}

So either $M = \bigcap_{S\in{\cal C}\cap {\cal A}}S$ is empty or a
cheater can be identified.

Next we will have a short look at the complexity of decision problems
related to conflict structures. If e.\,g. one player has doubts about the
validity of the assumption it is, in the worst case, difficult to test
if a conflict structure is consistent with the assumption that only
one set of $\cal A$ contains cheaters.

\begin{lemma}
For a given conflict structure $(\Gamma,{\cal A})$ deciding if the
assumption that only one set of $\cal A$ contains cheaters is
consistent with the graph of conflicts $\Gamma$ is
${\cal NP}$-complete.
\end{lemma}

\begin{proof}
If we set $\cal A$ to be the set of all subsets of $P$ with at most
$t$ players, then deciding consistency with a given conflict graph
$\Gamma$ is the same as deciding if there exists a vertex cover
with at most $t$ vertices. This is ${\cal NP}$-complete~\cite{GarJoh79}.
\end{proof}

In the worst case it is also difficult to identify a cheater by
deduction from the conflict structure.

\begin{lemma}
Identifying a
cheater by deduction from a given conflict structure $(\Gamma,{\cal A})$ is
${\cal NP}$-hard.
\end{lemma}

\begin{proof}
We show that a search algorithm which can identify a cheater whenever
a cheater can be identified by deduction from the conflict structure
can be used to decide if a graph has a unique vertex cover of
cardinality $t$.

We let $\cal A$ be the set of all subsets of $P$ with at most $t$
elements.
Let ${\tt identify}(\Gamma)$ be a search algorithm
which identifies a cheater if it is possible to identify a cheater.
We will use this algorithm to decide if there is a
unique solution to the vertex cover problem. As this uniqueness
problem is ${\cal NP}$-hard we then have shown the problem of
identifying cheaters to be ${\cal NP}$-hard
(see~\cite{Joh90} for the uniqueness problem and
\cite{GarJoh79} for the uniqueness preserving reduction of vertex
cover to the {\em satisfiability problem}).

To find a unique vertex cover (and hence decide its existence) we run
${\tt identify}(\Gamma)$ if no cheater is identified then there is not
a unique solution. If a cheater $p\in P$ is identified we
restrict our graph $\Gamma$ to $P\setminus \{ p\}$. We repeate this
procedure until ${\tt identify}$ has either found enough cheaters such
that they form a vertex cover, which then must be a unique vertex
cover, or not enough cheaters can be identified and no unique vertex
cover exists.
\end{proof}

A protocol for which it is necessary to identify cheaters whenever
possible can be impractical for large numbers of players. Fortunately
the protocols of this paper need to identify cheaters only in
situations where one player is in conflict with a set of players which
is not contained in $\cal A$. This player is hence easily identified
as a cheater. But we still have to be careful because with an
inappropriate presentation of the adversary structure $\cal A$ it can
even be difficult to decide membership in $\cal A$.

\begin{remark}
Let the adversary
structure $\cal A$ be given by sets $A_1,\dots,A_m$ such that ${\cal
A}= \{ A | \exists i\leq m:A\subseteq A_i\}$. Let the sets
$A_1,\dots,A_m$ be given by one boolean function $f(b_1,\dots,b_n)$
such that each input bit $b_i$ corresponds to a player $P_i\in P$ and a set
$A$ is in $\{A_1,\dots,A_m\}$ if the assigment 

$$b_i = \left\{\begin{array}{cc}
  1 & {\rm if\ }P_i\in A \\
  0 & else
  \end{array}\right.$$

is a satisfying assignment. Then deciding
membership in $\cal A$ is ${\cal NP}$-complete.
\end{remark}

\begin{proof}
Deciding membership in $\cal A$ for a set $A$ is clearly in $\cal NP$
as one can guess a superset of $A$ which yields a satisfying
assignemnet for $f$.

On the other hand we can reduce the satisfiability problem to deciding
membership in $\cal A$. As $\cal A$ contains for every set $A\in{\cal
A}$ all the subsets of $A$ the boolean function $f$ has a satisfying
truth assignment iff for one player $P_i$ we have $\{ P_i\}\in {\cal
A}$.
\end{proof}

\section{Committing to All Players}\label{SectionGBCX}

To ensure correctness of a multiparty protocol all players should be
committed to their inputs and to shares of intermediate results they
hold. Furthermore they should be able to give zero knowledge proofs
about properties of their inputs.

\subsection{Previous Results}

To be able to give zero knowledge proofs about properties of
commitments we use the following construction which can be found
in~\cite{CreGraTap95}.

\begin{definition}
A {\em bit commitment with Xor} ({\em BCX}) to a bit $b$ is a
commitment to bits $b_{1L}$, $b_{2L},\dots,$ $b_{mL},$ $b_{1R},\dots,$
$b_{mR}$
such that for each $i$ $b_{iL}\oplus b_{iR}=b$.
\end{definition}

The following result about zero knowledge proofs on BCX can as well be
found in~\cite{CreGraTap95} and in references therein.

\begin{theorem}\label{COPYworks}
Bit commitments with Xor allow zero knowledge proofs of linear
relations among several bits a player has committed to using
BCX. Especially (in)equality of bits or a bit string being contained
in a linear code.

Furthermore BCXs can be copied, as proofs may destroy a BCX.
\end{theorem}

\begin{proof}
We will not state a full proof here as it can be found
in~\cite{CreGraTap95}. But we will restate the copying procedure as it
is an important subprotocol of all of the following protocols.

Suppose Alice is committed to Bob to a bit $b$ and wants two instances
of this commitment. Then Alice creates $3m$ pairs of bit commitments
such that each pair Xors to $b$. Then Bob randomly partitions these
$3m$ pairs in three subsets of $m$ pairs, thus obtaining three
BCX and asks Alice to prove the equality of the first new BCX with
her BCX for $b$. This destroys the old BCX and one of the new BCX, but
an honest Alice can thereby convince an honest Bob that the two
remaining BCX both stand for the value $b$.
\end{proof}

Note that following this protocol of~\cite{CreGraTap95} it is possible
for a cheater to, with a polynomial probability, create an incorrect
BCX where a small (constant) number of pairs
$b_{iL},b_{iR}$ of plainly committed bits have an Xor unequal to the
bit committed by the BCX.  This does not harm the rest of the multiparty
computation as such a small inconsistency either has no influence on
the result or it is detected in the course of the protocol and leads
to a conflict as a zero knowledge proof or an unveil will be not
accepted (the same remark is necessary after Lemma~\ref{GBCXworks}).

In a multiparty scenario it is necessary that a player should be
committed to all other players.

\begin{definition}\label{DefGBCX}
A {\em global bit commitment with Xor} ({\em GBCX}) consists of BCX
commitments from one player from $P$ to a set of players which cannot
be a collusion such that
all players are convinced that this player did commit to the same bit
in all the different BCX.
\end{definition}

\begin{corollary}
Zero knowledge proofs of linear relations among several GBCX are
possible. Furthermore a GBCX can be copied by copying the individual
BCX.
\end{corollary}

For us it will be enough if a non-collusion (a set of players
trustable by definition) is convinced by the zero knowledge proof.

\subsection{Making Commitments More Robust}

We will use the GBCX protocol as it is presented in~\cite{CreGraTap95}
to bind a committing party to a set of players which cannot all
collude with the sender. Hence the bit cannot be changed by any allowed
collusion.

As the protocol for generating a GBCX needs coin tossing as a
subroutine we will briefly show that coin tossing is possible if no
two possible collusions cover the complete set of players.

\begin{remark}\label{CoinTossingWorks}
Given a set $P$ of $n$ players having access to a
broadcast channel and let every pair of players be connected by an
oblivious transfer channel, then $\cal A$-robust coin tossing is
possible if no two collusions of $\cal A$ cover $P$.
\end{remark}

\begin{proof}
Every player chooses a random bit and commits to it to every other
player. Then the bits are opened using the broadcast channel. Some
players might complain about other players.  Every bit accepted by a
non-collusion is called a valid bit. Then the result of the coin
tossing is chosen to be the Xor of the valid bits. Every player whose
bit is not accepted must be a cheater as he is in conflict with a non
collusion. As only a set of players contained in $\cal A$ can be
identified as cheaters and no two sets of $\cal A$ cover $P$ the bits
of a non-collusion will be accepted as valid. Therefore the resulting
bit is really random as it cannot be chosen by a collusion.
\end{proof}

It is easy to verify that after generating  a GBCX according
to~\cite{CreGraTap95} the sender of the commitments is bound to all
players who did not complain about the sender. Furthermore all players
who did not complain are convinced to hold commitments for the same
bit. For a given adversary structure $\cal A$ two cases can now occur:
\begin{enumerate}
\item A set $A\in{\cal A}$ of players complains about
the sender. Then the sender is bound to all players of the complement $A^c$  
of $A$ except to himself. The complete collusion necessary now to
change the bit would be $A^c$, which has to include the sender or the
sender would now complain about all other players and leave the protocol.
\item A set $A\not\in{\cal A}$ complains about the sender. Then the
sender has to leave the protocol.
\end{enumerate}

Whenever the sender is in conflict with a set $A$ of players then this
player can only change his commitment by colluding with all other
players of $A^c$ hence if no two possible collusions of $\cal A$ cover
the set $P$ of players then a player is either detected cheating or
his commitment is binding. The GBCX remains to be $2^P$-secure if used
this way.

Summarizing the above we can state the next result without further
proof. 

\begin{lemma}\label{GBCXworks}
For a set $P$ of players with each pair of players being connected by
an oblivious transfer channel and each player having access to a
broadcast channel and an adversary structure $\cal A$ for which
no two possible collusions cover $P$ it is possible to 
$\cal A$-partially robustly and $2^P$-securely
generate a GBCX or a cheater can be identified.
\end{lemma}

Note that following this protocol of~\cite{CreGraTap95} it is possible
for a cheater to, with a polynomial probability, create an incorrect
GBCX where a few users have a small (constant) number of pairs
$a^j_{iL},a^j_{iR}$ of committed bits which have an Xor unequal to the
bit committed to by the GBCX.  But this does not harm the rest of the
multiparty computation as such a small inconsistency either has no
influence on the result or it is detected in the course of the
protocol and leads to a conflict as a zero knowledge proof or an
unveil will be not accepted.

It is an interesting question if one could obtain a higher partial
correctness by sacrificing the $2^P$-securety, e.\,g., by exploiting
Lemma~\ref{OTworks} to obtain oblivious transfer between players in
conflict. But we will leave this question open in this paper.

\section{Distributed Bit Commitments}\label{SectionDBC}

Next we will consider the {\em distributed bit commitment}
of~\cite{CreGraTap95}. Such a {\em distributed bit commitment} consists of
several bit commitments each to a share of a bit. The multiparty
computation will later be computed on those shares.

\begin{definition}
A {\em distributed bit commitment (DBC) to a bit $b$} consists of $n$
GBCX one held by each player of $P$ such that the Xor of all values of
the individual GBCX equals $b$.

If the DBC is constructed in a way that one player knows how to unveil
all the GBCX the DBC consists of we say that it is a DBC of this player. 
\end{definition}

To create, according to~\cite{CreGraTap95}, a DBC of a player (Alice)
each player creates a GBCX to a random bit and opens it to Alice then
Alice creates a GBCX such that the Xor of all GBCX equals the bit $b$
she wants to commit to. The complete multiparty protocol will
perform circuit evaluation on the DBC of the players. The intermediate
results of this circuit evaluation will again be DBCs but for these no
player knows how to unveil all GBCX.

We will give a robust implementation of creating a DBC of a player in
our next result.

\begin{lemma}\label{DBCworks}
  Given an oblivious transfer channel between any two players and let
  every player have access to a broadcast channel, then
  for an adversary structure $\cal A$ which does not contain two sets
  covering all of $P$ an $\cal A$-partially robust multiparty protocol
  for creating a DBC of a player (Alice) can be implemented which is
  $2^P$-secure and if the protocol fails a cheater can be identified
  unambiguously
\end{lemma}

\begin{proof}
If Alice wants to generate a DBC for a bit $b$ all players have to
commit to a random bit using GBCX and then unveil this bit to
Alice. Then Alice will generate a GBCX such that the Xor of all the
bits equal the bit $b$.

The problem is that all the GBCX are only unveiled to Alice. Hence we
cannot distinguish between a party refusing to unveil to Alice and
Alice just claiming so. All other conflicts can be solved by
Lemma~\ref{GBCXworks}.

So assume Alice to be in conflict with a set of players $A$ while she
is creating a DBC. Then we will force the players from $A$ to unveil
their bits publicly. If some players are unable to unveil we have
identified cheaters.

We seem to loose a little bit
of security or correctness as the complement $A^c$ of the set $A$ can
reconstruct the secret. But as Alice is contained in $A^c$ the secret
can only be recovered if Alice is cheating, too. If Alice is part of
the collusion the collusion does not learn anything new by
reconstructing Alices input bit.
\end{proof}

\section{Committed Oblivious Transfer}\label{SectionGCOT}

Next we recall the definition of committed oblivious transfer, the key
protocol of~\cite{CreGraTap95}.

\begin{definition}
Given two players Alice and Bob where Alice is committed to bits
$a_0,a_1$ and Bob is committed to a bit $b$, then a {\em committed
oblivious transfer} protocol ({\em COT}) is a protocol where Alice
inputs information on her two commitments and Bob will input
data of his commitment and the result will be that Bob is committed
to $a_b$.

In a {\em global committed oblivious transfer} protocol ({\em GCOT})
all players are convinced of the validity of the commitments, i.e.,
that indeed Bob is committed to $a_b$ after the protocol.
\end{definition}

To achieve a robust version of this protocol (Lemma~\ref{GCOTworks}) we will
need an auxiliary protocol {\em forward oblivious transfer}
(Lemma~\ref{ForwardOT}) and a protocol which successfully implements
oblivious transfer even between players who are in conflict or a
cheater can be identified (Lemma~\ref{OTworks}).

\subsection{Forward Oblivious Transfer}

In this subsection we will introduce a protocol which allows a sender
(Alice) to implement an oblivious transfer to a receiver (Bob) she
is in conflict with. We will need
the help of a third player (Carol) who will learn all the data sent by
Alice, but will be unable to alter the data sent without getting in
conflict with either Alice or Bob. We call this protocol {\em forward
oblivious transfer} as the player Carol "forwards" the data to Bob
obliviously. 

\begin{lemma}\label{ForwardOT}
For three players Alice, Bob, and Carol where Carol is not in conflict
with Alice or Bob it is possible to implement a function {\bf Forward
Oblivious Transfer via} Carol {\bf of} $(a_0,a_1,b)$ where Alice
inputs two bits $a_0, a_1$, Bob inputs a bit $b$, Carol learns the two
bits $a_0, a_1$, and Bob learns only the bit $a_b$ for his choice of
$b$.  The protocol is $2^{\{Alice,Bob,Carol\}}$-partially
robust or a new conflict must arise.
\end{lemma}

\begin{proof}

We prove the claims of the lemma for the following protocol.

\vspace*{3mm}

\noindent {\bf Forward Oblivious Transfer via} Carol {\bf of}
$(a_0,a_1,b)$
\begin{enumerate}
\item Alice sends the bits $a_0, a_1$ to Carol.
\item Carol commits to $a_0,a_1$ to Alice and to Bob using a GBCX
involving only the players Alice, Bob, and Carol. Then Carol opens the
commitment to Alice to convince her that she is now committed to
$a_0,a_1$ to Bob.
\item Bob commits to a bit $b$ to Carol.
\item Carol runs COT($a_0,a_1,b$) with Bob.
\end{enumerate}

\vspace*{3mm}

For the security of the protocol we have to prove that 
\begin{enumerate}
\item Alice and Carol cannot together learn the secret
$b$ of Bob. 
\item Bob alone cannot learn the secret $a_0,a_1$ of Alice
(together with Alice or together with Carol $a_0,a_1$ are not secret
any more as they can be derived from the input resp. output of the function.) 
\end{enumerate}
Point 1. is clear from the security of the COT protocol. Point
2. follows directly from the security of the GBCX protocol and the COT
protocol.

To prove the partial correctness it is enough to prove that Caro alone
cannot alter the two bits without getting in conflict with Alice or Bob.
Alice can check if the two bits Carol is committed to equal the bits
she sent to Carol because of the binding property of the GBCX bit
commitment. Bob can check if the bits Carol is committed to equal the
bits Alice sent to Carol by the properties of the COT protocol used.
\end{proof}

\subsection{GCOT from Forward Oblivious Transfer}

The player helping in the protocol forward oblivious transfer
learns all bits transmitted. To keep up the security we will use the
protocol many times with different helpers to obtain oblivious
transfer even between players in conflict. Then the secret is
distributed among all helping players.

\begin{lemma}\label{OTworks}
  Given an oblivious transfer channel between any two players as well
  as a broadcast channel, then for an adversary
  structure $\cal A$ for which no two sets cover $P\setminus \{P_i\}$ for
  any player $P_i$ an $\cal A$-partially robust, $\{ A\subseteq P |
  A^c\not\in{\cal A}\}$-secure multiparty protocol for oblivious
  transfer can be implemented such that the sender is committed to
  what he sent and whenever a party complains about the result of
  the protocol a new conflict arises.
\end{lemma}

\begin{proof}
If the sender and the receiver of an oblivious transfer are not in
conflict yet, then a new conflict arises as soon as one party
complains. So we are left with the interesting case where the sender
and the receiver are already in conflict. In this situation we use the
following protocol:

\noindent {\bf Oblivious Transfer for players in conflict}($a_0,a_1,b$)
Let $M$ be the set of players not in conflict with Alice or Bob.
\begin{enumerate}
\item Bob chooses a bit $b$
\item For all $p\in M$ do
\begin{enumerate}
\item Alice chooses random bits $a_{0,p},a_{1,p}$ and performs with
Bob {\bf
Forward Oblivious Transfer via} $p$ {\bf of} $(a_{0,p},a_{1,p},b)$
\item If Alice or Bob gets in conflict with $p$ then let $M:= M\setminus\{p\}$
\end{enumerate}
\item Alice calculates $a_0 \oplus \bigoplus_{p\in M} a_{0,p}$ and 
$a_1 \oplus \bigoplus_{p\in M} a_{1,p}$ and broadcasts these two
bits. 
\end{enumerate}

We now prove the security, partial correctness, and fairness of the
above protocol.

Security: The secret bit $b$ of Bob cannot be learnt by anyone due to
the security of the COT protocol.  Now we look at Alices secrets.  Let
$B$ denote the set of players the receiver Bob is in conflict with and
$A$ be the set the sender Alice is in conflict with.  The players of
the set $M$ can together reconstruct a secret of the sender Alice. But
the set $M$ cannot contain all cheaters, the complete collusion is
larger. If Alice is honest (otherwise we don't need to protect her
secret), then all players of $A$ are cheaters and have to be
considered as part of the collusion. The complete collusion able to
reconstruct a secret bit and containing all cheaters is then at least
as large as $A\cup M = B^c$. The set $B$ is contained in $\cal A$,
otherwise Bob would have left the protocol, then $B^c\not\in{\cal
A}$ and no collusion of $\cal A$ learns a secret. It remains to be
shown that no honest but curiuous player gets to know a secret. As
$|M|>1$, because no two collusions cover all but one player, Alices
secret is always distributed among several honest players and no
single honest but curious player can reconstruct it. We can conclude
that the protocol is $\cal A$-secure.

Partial correctness: According to Lemma~\ref{ForwardOT} no player of
$M$ can have altered the values of the bits without a new conflict
arising. At the end of the protocol the set $M$ contains only the
players Alice and Bob are not in conflict with. Thus the players of
$M$ cannot have altered the bits, hence we even get $2^P$-partial
correctness for this protocol.

Fairness is not an issue here as only one player, Bob, learns a result.

The sender is committed to bits $a_0,a_1$ as each player $p\in M$
is committed to the bits $a_{0,p}, a_{1,p}$ the sender can ask all
players from $M$ to open the bits. If the bits are not opened
correctly either the sender or the receiver will object and a new
conflict must arise between a player from $M$ and Alice or Bob.

\end{proof}

Our next result will show that all steps of the GCOT protocol
of~\cite{CreGraTap95} can be verified by other players except one step
involving an oblivious transfer between two players. If a conflict
arises in this step we can replace the oblivious transfer by the
protocol of  Lemma~\ref{OTworks}.

\begin{lemma}\label{GCOTworks}
Let $P$ be a set of players where each pair of players is connected by
an oblivious transfer channel and every player has access to a
broadcast channel. Let $\cal A$ be an adversary structure for which no
two collusions cover $P\setminus \{P_i\}$ for any player $P_i$. Then a
GCOT protocol can $\cal A$-partially robustly and $\{ A\subseteq P |
A^c\not\in{\cal A}\}$-securely be implemented between two players who
are in conflict or a cheater can be identified.
\end{lemma}

\begin{proof}
We will restate the GCOT protocol of~\cite{CreGraTap95} without a
proof of its security. Details can be found in~\cite{CreGraTap95}.
Then we will carefully investigate the steps and see, that by
replacing GBCX with the modified protocol of Lemma~\ref{GBCXworks} and
using the oblivious transfer of Lemma~\ref{OTworks} each step either
works, or a new conflict arises, or a cheater is identified. The steps
which did not work can be repeated and eventually the protocol works
or a cheater can be identified unambiguously. In the restated protocol
we will use the notation of~\cite{CreGraTap95}: indices are
superscript and ${\rm OT}(a_0,a_1)(b)$ denotes the {\bf Oblivious
Transfer for players in conflict}($a_0,a_1,b$) protocol of
Lemma~\ref{OTworks}.

\vspace*{3mm}

{\bf GCOT}$(a_0,a_1)(b)$
{\small
\begin{enumerate}
\item All participants together choose one decodable $[m,k,d]$ linear
code $\cal C$ with $k>(1/2+2\sigma)m$ and $d>\epsilon n$ for positive
constants $\sigma,\epsilon$, efficiently
decoding $t$ errors.
\item Alice randomly picks $c_0,c_1\in{\cal C}$, commits to the bits
$c_0^i$ and $c_1^i$ ($i\in \{1,\dots,m\}$) of the code words, and
proves that the codewords fulfil the linear relations of $\cal C$.
\item Bob  randomly picks $I_0,I_1\subset \{1,\dots,M\}$, with
$|I_0|=|I_1| = \sigma m,$ $I_1\cap I_0 =\emptyset$ and sets
$b^i\leftarrow \overline b$ for $i\in I_0$ and $b^i \leftarrow b$ for
$i\not\in I_0$.
\item Alice runs ${\rm OT}(c_0^i,c_1^i)(b^i)$ with Bob who gets $w^i$
for $i\in \{1,\dots,m\}$.
Bob tells $I=I_0\cup I_1$ to Alice who opens  $c_0^i,c_1^i$ for each
$i\in I$.
\item Bob checks that $w^i = c_{\overline b}^i$ for $i\in I_0$ and
$w^i = c_{b}^i$ for $i\in I_1$, sets $w^i\leftarrow c_b^i$, for $i\in
I_0$ and corrects $w$ using $\cal C$'s decoding algorithm, commits to
$w^i$ for $i\in \{1,\dots,m \}$, and proves that $w^1\dots w^m\in
{\cal C}$.
\item All players together randomly pick a subset $I_2\subset
\{1,\dots,m\}$ with $|I_2|=\sigma m$, $I_2\cap I=\emptyset$ and Alice
opens $c_0^i$ and $c_1^i$ for $i\in I_2$.
\item Bob proves that $w^i = c_b^i$ for $i\in I_2$.
\item Alice randomly picks and announces a privacy amplification
function $h:\{0,1\}^m\rightarrow \{0,1\}$ such that $a_0 = h(c_0)$ and
$a_1 = h(c_1)$ and proves $a_0= h(c_0^1,\dots,c_0^m)$ and $a_1=
h(c_1^1,\dots,c_1^m)$.
\item Bob sets $a\leftarrow h(w)$, commits to $a$ and proves $a =
h(w^1\dots,w^m)$.
\end{enumerate}
} 
As GBCX commitments as well as zero knowledge proofs convincing a
non collusion can be performed
by all players unless a cheater is identified (Lemma~\ref{GBCXworks})
the honest behaviour of Alice and Bob can be checked by a non
collusion in
all steps, but in step 4.

If now Bob claims that Alice cheated in step 4. then Alice can open
the codewords $c_0,c_1$ according to Lemma~\ref{OTworks} then either
Alice or Bob are caught cheating or if the opening was not successful
a new conflict must arise (Lemma~\ref{OTworks}). If this is the case
we repeat the steps 1. to 4. with new random choices. After a finite
number of repetitions a cheater will be identified as there cannot be
arbitrarily many conflicts. As the codewords which might have to be
opened are random and not related to Alices secret inputs no security
is lost by restarting the protocol. Hence the security is the same as
stated in Lemma~\ref{OTworks}.
\end{proof}

\section{Circuit Evaluation on DBCs}\label{SectionProtocol}

In the previous sections we developed enough tools to now state the
complete protocol which very closely follows the protocol
of~\cite{CreGraTap95}, but uses the more robust protocols for GBCX,
DBC, and GCOT introduced so far. For the convenience of the reader we
restate those results and proofs of~\cite{CreGraTap95} needed to
picture the complete protocol.

First we restate the definition of the boolean function AND on
commitments as we will use it for the multiparty protocols later.

\begin{definition}
A {\em pair and} ({\em PAND}) is a protocol which takes as input two
BCX one from a player Alice and one from a player Bob and outputs two
BCX one for Alice and one for Bob such that the Xor of the values of
the new BCX equal the AND of the values of the input BCX.

A {\em global pair and} ({\em GPAND}) is a generalization of PAND to a
set of players. Two active players (Alice and Bob) perform a PAND in a
way that all other players are convinced of the Xor of the new
commitments equals the AND of the input values.

By {\em and} ({\em AND}) we will denote a protocol which takes as
input two DBC and outputs one DBC representing the AND of the
values of the input DBCs such that every party is convinced of this.
\end{definition}

In~\cite{CreGraTap95} it is shown how to obtain an AND on DBCs from a
protocol for GCOT: 

\begin{lemma}\label{ANDworks}
  With the notation of Lemma~\ref{OTworks} we have:
  Given an oblivious transfer channel between any two
  players and a broadcast channel then an
  $\cal A$-partially robust and
  $\{ A\subseteq P | A^c\not\in{\cal A}\}$-secure
  multiparty protocols for GPAND and AND can be implemented such
  that whenever a party complains about the result of the protocol
  a cheater is identified.
\end{lemma}

\begin{proof}
We restate the protocols from~\cite{CreGraTap95} to see
that they involve only primitives which can be dealt with according to
our results so far.
 
A PAND can be realized by the following protocol: 
Alice is committed to $a$ and Bob is committed to $b$. Then Alice
chooses a random bit $a'$ and runs COT$(a', a'\oplus a)(b)$ with
Bob who gets $b'$. We have $a'\oplus b' = a\wedge b$ because for $b =
0$ we have $b' = a'$ and hence $a'\oplus b'=0$, for $b=1$ we get $b'=
a\oplus a'$ and $a'\oplus b'=a$.

For a GPAND protocol the COT protocol has to be replaced by GCOT.

To evaluate an AND on DBCs we observe that $$(\bigoplus_{i=1}^n a_i)\wedge
(\bigoplus_{j=1}^n b_j) = \bigoplus_{i,j=1}^n (a_i\wedge b_j).$$
From this we can conclude that an AND operation on DBCs can be realized
by $n^2$ GPAND one for each pair of players and Xor operations for
each player.
\end{proof}

To be able to make circuit evaluation for all possible boolean
functions we also need a NOT on DBCs.

\begin{remark}\label{NOTworks}
Given a set $P$ of players, a DBC of these player, and an adversary
structure $\cal A$ for which no two sets of $\cal A$ cover $P$, then
there exists a protocol which is $\cal A$-partially robust,
$2^P$-secure, and
successfully inverts the bit the DBC stands for or a cheater is
identified.
\end{remark}

\begin{proof}
To implement such a NOT gate one player is picked who must invert his
GBCX (his ``share'' of the DBC which represents a bit $b$). The player
generates a new GBCX and proves that it is unequal to the GBCX he held
before. This GBCX together with the GBCX of the other players form a
DBC for the inverted bit.
\end{proof}

All protocols presented so far are only $\cal A$-partially correct,
but they allow the identification of a cheater if they fail. To obtain
$\cal A$-correct protocols from these we use a very simple idea, we
will restart the protocol every time it failed without the players who
have been caught cheating.  The exclusion of cheating players can
change the value of the function to be computed. The best solution
to this problem would be to have a default input like ``unvalid''. But
the effect of the exclusion of cheating players does not affect the
correctness of the protocol as a cheating player
could as well have chosen a nonsensical input. In Remark~\ref{Restart}
we will deal with the problem that some players might try to change
their inputs after a restart.

With the protocols presented so far and restarting the protocol if it
fails we get:

\begin{lemma}\label{ProtocolSketch}
  Using the notation of Lemma~\ref{OTworks} we get:
  Given an oblivious transfer channel between any two
  players as well as a broadcast channel, then every
  function can be implemented by a multiparty protocol which is $\cal
  A$-robust and $\widetilde{\cal A}$-secure if the following conditions
  hold:
\begin{enumerate}
\item the adversary structure $\cal A$ does not contain two sets
  covering $P\setminus \{P_i\}$ for any $P_i\in P$ and
\item the adversary structure  $\widetilde{\cal A}$ does not contain a
  complement of a set of $\cal A$.
\end{enumerate}
\end{lemma}

\begin{proof}
According to Lemma~\ref{ANDworks} and Remark~\ref{NOTworks} we can
realize the boolean operation AND and NOT on DBCs such that whenever
the protocol fails a cheater is identified. Furthermore we can
generate DBCs successfully or a cheater will be
identified (Lemma~\ref{DBCworks}). Using these techniques we will implement
oblivious circuit evaluation. The protocol will be restarted each time
it had to be aborted, but without the players which were identified as
cheaters. 

We will next have to clarify how a protocol begins and how it is
ended. Below we will sketch the structure of the comlete protocol,
without mentioning possible restarts, closely
following~\cite{CreGraTap95}.

{\bf Initialization Phase}: All players have to agree on the function
to be computed as well as on the circuit $F$ to be used, they have to
agree on an adversary structure $\cal A$ such that the protocol will
be $\cal A$ robust and all players have to agree on the security
parameters used and on a code $\cal C$ for the GCOT protocol.
Furthermore the players agree on how to, in case of a restart of the
protocol, choose the input of a cheater which has been excluded from
the protocol.

Then all players create DBCs to commit to their inputs.

{\bf Computing Phase}: The circuit is evaluated using AND and NOT
gates on the input DBCs. If the circuit requires 
several copies of a DBC then a DBC is copied by
copying the GBCX it consists of. A GBCX can be copied by copying all
its BCX with the procedure of Theorem~\ref{COPYworks}. 

{\bf Revelation Phase}: The result of a computation is hidden in
DBCs. These have to be unveiled in a way to ensure the fairness of the
protocol. Following~\cite{CreGraTap95} we use the techniques
from~\cite{Cle89,GolLev90} to fairly unveil the secret information
such that no collusion can run off with an advantage of more than a
fraction of a bit.  Of course an $\widetilde{\cal A}$-secure protocol
cannot be more than $\widetilde{\cal A}$-fair.
\end{proof}

\section{Higher Security by a More Careful Analysis}

The result of Lemma~\ref{ProtocolSketch} is a little bit too
pessimistic. It does not take into account that the GCOT protocol has
to work only in one direction between every pair of
players. Exploiting this property we will be able to obtain
security against one more collusion $B$ which may be a complement of a
set of the adversary structure $\cal A$.

We first take a closer look at the situation when a complement of
a set from $\cal A$ contains all cheaters and is able to
reconstruct a secret bit:

\begin{remark}\label{Complement}
If for an $\cal A$-robust protocol implemented according to
Lemma~\ref{ProtocolSketch} there exists a set $B\in\cal A$ such that
its complement $B^c\not= P$ contains all cheaters and is able to reconstruct a secret bit
which cannot be reconstructed from the input of the players from $B^c$
and the output of the protocol then all of the following conditions
hold:
\begin{enumerate}
\item Lemma~\ref{OTworks} was used to realize oblivious transfer
between two players. 
\item The receiver of this oblivious transfer is in conflict with all
players from a set containing $B$ and is not in conflict with any
player who is in conflict with the sender.
\item The sender of this oblivious transfer is honest and the receiver
is cheating.
\end{enumerate}
\end{remark}

\begin{proof}
By inspection of the Lemmata~\ref{ProtocolSketch}, \ref{ANDworks},
\ref{OTworks} we can see, that the only step where the $2^P$-security
is lost is the use of Lemma~\ref{OTworks}. The secret which is
distributed when applying Lemma~\ref{OTworks} is a secret of the
sender in the oblivious transfer by Lemma~\ref{OTworks}. Hence the
$2^P$-security is lost only if the sender was honest. To complete the
proof we look at the set which can reconstruct the distributed secret.

Let a player (Alice) be in in conflict with a set $A$ containing a
player Bob and Bob being in conflict with a superset $C$ of the set
$B$.  Of course $C$ contains Alice. If we use Lemma~\ref{OTworks} to
implement oblivious transfer between Alice and Bob then secret bits of
Alice are distributed among the players of $P\setminus(A\cup C)$ a
subset of $P\setminus(A\cup B)$. If Alice is honest (otherwise we need
not protect her secret) then all players of $A$ are cheating and the
complete collusion able to reconstruct secret bits of Alice is
$A\cup P\setminus(A\cup C) = (C\setminus A)^c$ a subset of
$(B\setminus A)^c$. The set $(C\setminus A)^c$ contains all cheaters
and can reconstruct a secret of Alice, but it can only be a subset of
$B^c$ if $A$ and $B$ are disjoint, i.\,e., if $B$ does not contain any
player in conflict with the sender. 
\end{proof}



From the proof of Lemma~\ref{ProtocolSketch} and Lemma~\ref{ANDworks} we can
see that the GCOT within the AND protocol has to work only in one
direction between every pair of players.  Using this simple
observation together with the above remark we are ready to state the
main result of this section.

\begin{lemma}\label{AprioriLemma}
Let $P$ be a set of $n$ players with every pair of players being
connected by an oblivious transfer channel and every
player having access to a broadcast channel. Let $\cal A$ and
$\widetilde{\cal A}$ be adversary structures, then for all functions
$\cal A$-robust and $\widetilde{\cal A}$-secure multiparty protocols exist
if
\begin{enumerate}
\item the adversary structure $\cal A$ does not contain two sets
  covering $P\setminus \{P_i\}$ for any $P_i\in P$ and
\item the adversary structure $\widetilde{\cal A}$ contains only the
complement of one previously chosen set $B$ which is maximal in $\cal A$.
\end{enumerate}
\end{lemma}

\begin{proof}
Let $B$ be any maximal set of $\cal A$.  In addition to
Lemma~\ref{ProtocolSketch} we have to prove that we can additionally
prevent $B^c$ from reconstructing any secret data. 

From Lemma~\ref{Complement} we know that a complement of a maximal set
$B$ contains all cheaters and can reconstruct a secret only if the
receiver of an oblivious transfer by Lemma~\ref{OTworks} was in
conflict with a superset of $B$. As $B$ is maximal either the receiver
is detected cheating by being in conflict with a set not in $\cal A$
or the receiver has to be in conflict with exactly all players from
$B$.  We keep in mind that oblivious transfer, as well as GCOT, is
needed in one direction only between every pair of players.  We modify
the protocol such that a player who is in conflict with exactly the
players of $B$ always sends in an oblivious transfer if it is
implemented by Lemma~\ref{OTworks}. It remains to be shown that it is
impossible that the receiver and the sender are in conflict with the
players of $B$. Lemma~\ref{OTworks} is only employed if the sender
and the receiver are in conflict. Hence the the sets of players the
sender and the receiver are in conflict with have to differ as no one
can be in conflict with himself.   
\end{proof}

In the above result one can see the trade off between robustness and
security. The smaller $\cal A$ can be chosen the larger $\widetilde{\cal
A}$ will be.

\section{The Security of the Protocol After
Termination}\label{SectionAfterTermination} 

The result of Lemma~\ref{AprioriLemma} guarantees us $\cal
A$-robustness and $\{ A\subseteq P | A^c\not\in{\cal A}\}\cup \{
B^c\}$-security for a previously chosen $B\in{\cal A}$,
but the security can be even higher depending on the course of the
protocol. A trivial example is that $2^P$-security is achieved if no
player complained during the protocol, because in this case the
protocol specializes to the protocol of~\cite{CreGraTap95}.

In this section we want to derive the security the protocol guarantees
from the knowledge one has after termination. We will see that the
the security will be higher than guaranteed by
Lemma~\ref{AprioriLemma}.

To clearly distinguish the security guaranteed in advance and the
security which is actually obtained we will speak of {\em a priory
security} and {\em a posteriori security}.

\begin{remark}\label{onlytwo}
Whenever no cheater can be identified and two players (Alice and Bob)
are in conflict with two disjoint maximal sets $A$, $B$ of $\cal A$
respectively, then every other conflict present must be a conflict
between a player from the set $A$ and a player from the set $B$.
\end{remark}

\begin{proof}
Let Alice be in conflict with the set $A$ maximal in $\cal A$ and Bob
be in conflict with the set $B$ maximal in $\cal A$ and let $A\cap
B=\emptyset$. Alice and Bob must be in conflict, because only one
collusion from $\cal A$ is cheating. Hence Alice $\in B$ and Bob $\in
A$. We now look at any additional conflict. This conflict has to
involve a player from $A$ otherwise we could identify a cheater,
because $A$ would not be a vertex cover of the conflict graph and Bob
must be cheating as he is in conflict with all players of $A$. For the
same reason one of the two players in conflict must be contained in
$B$ else Alice would be caught cheating. Hence every additional
conflict is a  conflict
between a player from the set $A$ and a player from the set $B$.
\end{proof}

With this remark and further exploiting the fact that oblivious
transfer is needed only in one direction between every pair of players
we get the main result of this paper:

\begin{theorem}\label{MPwithOT}
Let $P$ be a set of $n$ players with every pair of players being
connected by an oblivious transfer channel and every player having
access to a broadcast channel. Let $\cal A$, $\widetilde{\cal A}$, and
$\widehat{\cal A}$
be adversary structures, then for all functions $\cal A$-robust
multiparty protocols exist which are a priori $\widetilde{\cal A}$-secure
and a posteriori $\widehat{\cal A}$-secure
if
\begin{enumerate}
\item the adversary structure $\cal A$ does not contain two sets
  covering $P\setminus \{P_i\}$ for any $P_i\in P$,
\item for a previously chosen set $B$ maximal in $\cal A$
the adversary structure $\widetilde{\cal A}$ does not contain
any complement of a set of $\cal A$ except $B^c$, and
\item there exists a set $A$ $(\not= B)$ maximal in $\cal A$ such that the
adversary structure $\widehat{\cal A}$ does not contain
any complement of a set of $\cal A$ except the sets of $\{S\in{\cal
A}| S {\rm \ is\ maximal\ and\ } S\not=A \}$.
\end{enumerate}
\end{theorem}

\begin{proof}
All properties of $\cal A$ and $\widetilde{\cal A}$ were dealt with in
Lemma~\ref{AprioriLemma}. Hence we will have
to consider only the a posteriori security in this proof.


To achieve the security stated in point 3. of the theorem we need one
more modification of the protocol developed so far. Again we keep in
mind that oblivious transfer has to be used in one direction only
between every pair of players. We will introduce more rules regulating
the direction in which oblivious transfer has to be used whenever
Lemma~\ref{OTworks} is employed.

\begin{enumerate}
\item If a player is in conflict with the set $B$ then this player
sends only in an oblivious transfer which is implemented by
Lemma~\ref{OTworks}.
\item If a player is in conflict with a maximal set of $\cal A$ and
needs to employ Lemma~\ref{OTworks}, then this
player always sends to players who are not in conflict with a maximal
set of $\cal A$. 
\item
If two players are in conflict with a maximal set of $\cal A$ then we
use a previously fixed order $<$ on the set of maximal sets of $\cal
A$. The player in conflict with the maximal set larger with respect to
the order $<$ sends and the player in conflict with the maximal set
smaller with respect to the order $<$ receives. To be consistent with
the above the set $B$ must be maximal with respect to the ordering $<$. 
\end{enumerate}

These additional rules for the direction of oblivious transfers are in
accordance with the proof of Lemma~\ref{AprioriLemma}. Hence we don't
need to prove points 1. and 2. from the above theorem as the proof of
Lemma~\ref{AprioriLemma} still applies.

Now we prove point 3. of the above theorem.

A set $A^c$ with $A$ maximal in $\cal A$ is able to reconstruct a
secret only if the receiver of an oblivious transfer was in conflict
with the players of $A$ and $A$ does not contain any players the
sender is in conflict with (Remark~\ref{Complement}). Let the sender
be in conflict with the players of a set $A'$ then $A'$ must be
maximal in $\cal A$ or the direction of the oblivious transfer would
have to be different (point 2. of the above enumeration). Furthermore
$A$ and $A'$ must be disjoint as otherwise $A$ would contain a player
the sender is in conflict with. We can conclude that a set $A^c$ with $A$ maximal in $\cal A$ is able to reconstruct a
secret only if there exist two players each of which is in conflict
with a maximal set and these two sets are disjoint.
From now on we will consider only this situatiuon. From
Lemma~\ref{onlytwo} we know that in such a situation every conflict is
a conflict of a player from $A$ and a player of $A'$. So whenever in
this case two
players are in conflict with maximal sets of $\cal A$ one set must be 
$A$ and the other must be $A'$.

During an oblivious transfer which is implemented by
Lemma~\ref{OTworks} only $A^c$ or $A'^c$ could learn a secret bit, but
according to point 3. of the above enumeration the direction of the
oblivious transfer is always chosen in a way that only among the
players of one of these
two sets secrets will be shared. So from the complements of the sets
which are maximal in $\cal A$ only one set is excluded from
$\widehat{\cal A}$ all other sets of $\widehat{\cal A}$
were already contained in $\widetilde{\cal A}$. 
\end{proof}

Note that we cannot choose the adversary structure $\widehat{\cal A}$ as
we cannot choose who will be in conflict with whom. But it is obvious
that the  adversary structure $\widehat{\cal A}$ can be read off the set
of all conflicts which did occur during the oblivious transfer steps
of the protocol. So  $\widehat{\cal A}$ is known after the protocol
terminated.

Even after termination of a multiparty protocol there is a difference
between partial robustness and robustness. After termination of a
protocol no collusion can change the result or abort the protocol
anymore hence partial robustness is equivalent to security then. But
robustness additionally requires that no single honest but curious
player learns a relevant secret if a collusion of players leaks their
secret data.

We will see in the following that there is a very easy relation
between security and robustness after termination of a protocol. This
allows us to describe the robustness our protocols achieve during
their execution and after termination by Theorem~\ref{MPwithOT} and
the following lemma.

\begin{lemma}
A multi party protocol which is $\cal A$-secure after termination is
$\cal B$-robust after termination for ${\cal B} = \{ B | \exists
A\in{\cal A}:B\subset A {\rm \ and\ } B\not= A \}$.
\end{lemma}

\begin{proof}
After termination of the protocol no collusion can change the result
or abort the protocol anymore hence the only problem left is that a
collusion of players leaks their secrets. If a collusion leaks their
secret data it can happen that an honest but curious player learns a
secret without himself colluding. But it is obvious that for an $\cal
A$-secure protocol the collusion which is leaking secrets 
plus the honest but curious player must not be contained in $\cal A$
if the honest but curious player is to learn a secret. Hence after
termination the protocol is $\cal B$-robust for ${\cal B} = \{ B | \exists
A\in{\cal A}:B\subset A {\rm \ and\ } B\not= A \}$.
\end{proof}

Like the protocols of~\cite{CreGraTap95} our protocols are efficient
as none of our additional protocol steps needs non polynomial
resources.

\begin{corollary}
The protocols of Theorem~\ref{MPwithOT} are efficient in the number
$n$ of players and the size of the circuit used to evaluate the
function to be computed.
\end{corollary}

One problem is left. If the inputs of the multiparty computation are
time critical we have to keep the players from changing their inputs
if the protocol has to be restarted.

\begin{remark}\label{Restart}
To avoid that players change their inputs in a restarted protocol one
can let every player commit to his inputs by GBCX before the protocol
starts. When restarting the protocol the players have to prove the
equality of their newly generated DBC and their original inputs.

As the inputs could even depend on the conflicts appearing we don't
allow any complaints in the commit phase before the protocol until
all players claim to have committed to all other players.  
\end{remark}

Following this remark it is clear that every player which will not be
caught cheating had to commit his input to a non collusion. This is
enough to ensure that every player who will not be expelled from the
protocol cannot change the input he committed to following the above
remark.

%
%

\section{An Application to Quantum Multiparty Protocols}\label{QMP}

The main problem with quantum protocols like bit commitment is that
measurements can be delayed thereby allowing one party to
cheat~\cite{May96,LoCha96}.  In~\cite{ImaMue00QMPforIEICE} secret
sharing is used to force honest measurements, following ideas
from~\cite{Cre94,Yao95}. Once these measurements are performed the
assumptions about possible collusions can be loosened. This kind of
temporary assumptions are a very interesting feature of quantum
cryptographic protocols. In~\cite{ImaMue00QMPforIEICE} the following
result is proven:

\begin{corollary}\label{CorOTwithQC}
Let $P$ be a set of players for which each pair of players is
connected by a quantum channel and an authenticated insecure channel
and every player has access to a broadcast channel. Let $\cal A$ be an
adversary structure for which no two collusions cover the set
$P\setminus \{P_i\}$ for any player $P_i$ and let $M$ be any maximal
set in $\cal A$. Then for every pair of players, who will not be in
conflict after the protocol, an oblivious transfer is possible from one
of the two players to the other player which is $\cal A$-robust and $\{A^c
| A\not\in {\cal A} \}\cup M^c$-secure.
\end{corollary}

The problem that oblivious transfer is only implementable between
players not in conflict with each other can be solved by the protocols
of this paper as normal oblivious transfer cannot be used sometimes as
well if players are in conflict.  With the results of this paper we
are able to give a full proof of the main result
of~\cite{ImaMue00QMPforIEICE}.

\begin{theorem}\label{MainResult2}
  Let $P$ be a set of players each having access to a broadcast
  channel and every pair of players of $P$ being connected by a
  quantum channel and an insecure but authenticated classical
  channel. 
  Then $\cal A$-robust  quantum multiparty protocols 
  for all functions exist if and only if no two collusions of $\cal A$
  cover $P\setminus\{ P_i\}$ for any player $P_i$.

  These protocols are $\widetilde {\cal A}$-secure after termination
  if and only if the adversary structure $\widetilde{\cal A}$ contains
  at most one complement of a previously chosen set from ${\cal A}$.
\end{theorem}

%
%

%
%

\section{Conclusions and Outlook}

We presented multiparty protocols with oblivious transfer which can
tolerate disruptors if no two possible collusions of active cheaters
together contain all but one players. This is optimal.

If the set of all possible collusions of active cheaters is denoted by
$\cal A$ then our protocols are $\widetilde{\cal A}$-secure for an
$\widetilde{\cal A}$ which does not contain only one complement of a
set of $\cal A$. After termination the protocol is $\widehat{\cal A}$-secure 
for an adversary structure $\widehat{\cal A}$ which does not contain
any complement of a set of $\cal A$ except the sets of $\{S\in{\cal
A}| S {\rm \ is\ maximal\ and\ } S\not=A \}$.

Our result has implications on quantum cryptography maybe opening new
kinds of applications to quantum channels.

We conjecture that the security of the presented protocols cannot be
substantially improved. But the problems dealt with in this paper
indicate that there might be a more powerful cryptographic primitive
than oblivious transfer (we recently learnt about independent research
in this direction~\cite{FitGarMauOst00}). Every cheater has to take
care not to get into conflict with too many players, because the
cheater could then be identified. A more powerful primitive could make
it difficult to know whom one is cheating.  A candidate could be an
{\em anonymous oblivious transfer} where the receiver is picked at
random. Whenever a player tries to send trash over an anonymous
oblivious transfer channel this player risks to be in conflict with
the receiver which is picked at random.

Another interesting question is whether such an anonymous oblivious transfer
channel could be implemented---relative to reasonable assumptions---by
a quantum channel. This would imply that there exist situations in
which a quantum channel is cryptographically more powerful than
oblivious transfer.


%
%


\end{document}